# The Ebb and Flow of Brand Loyalty: A 28-Year Bibliometric and Content Analysis

**Azin Yazdi,** Department of Management, Payame Noor University, Iran.
Email: azinyazdi36@gmail.com

**Sunder Ramachandran,** Abu Dhabi School of Management, Abu Dhabi, United Arab Emirates.
Email: sunderrama@gmail.com

**Hoda Mohsenifard,** Islamic Azad University, Bushehr Branch, Iran. Email:
hoda.mohsenifard@gmail.com

**Khaled Nawaser,** Arvandan Non-Profit Higher Education Institute, Khorramshahr, Iran. Email:
khalednawaser56@gmail.com

**Faraz Sasani,** School of Business and Economics, Humboldt University of Berlin, Berlin, Germany. Email: sasanifaraz@gmail.com

**Behrooz Gharleghi**, Arden University, Berlin, Germany. Email: bgharleghi@arden.ac.uk

## Abstract

Business research is facing the challenge of scattered knowledge, particularly in the realm of brand loyalty (BL). Although literature reviews on BL exist, they predominantly concentrate on the present state, neglecting future trends. Therefore, a comprehensive review is imperative to ascertain emerging trends in BL This study employs a bibliometric approach, analyzing 1,468 papers from the Scopus database. Various tools including R software, VOS viewer software, and Publish or Perish are utilized. The aim is to portray the knowledge map, explore the publication years, identify the top authors and their co-occurrence, reliable documents, institutions, subjects, research hotspots, and pioneering countries and universities in the study of BL. The qualitative section of this research identifies gaps and emerging trends in BL through Word Cloud charts, word growth analysis, and a review of highly cited articles from the past four years. Results showed that highly cited articles mention topics such as brand love, consumer-brand identification, and social networks and the U.S. had the most productions in this field. Besides, most citations were related to Keller with 1,173 citations. Furthermore, in the qualitative section, social networks and brand experiences were found to be of interest to researchers in the field. Finally, by introducing the antecedents and consequences of BL, the gaps and emerging trends in BL were identified, so as to present the direction of future research in this area.

## Introduction

The domain of consumer behavior has garnered considerable attention in marketing, with a plethora of studies conducted in this area (Cohen et al., 2014; Jahanshahi et al., 2020). This is due to the recognition that theoretical insights into consumer behavior are instrumental in developing effective marketing strategies (Abrudan, et al., 2020). The nexus between consumer behavior and brands is salient in extant literature (Batra, 2019). In the context of marketing, brands serve as instruments to distinguish products by endowing them with value. In contemporary marketing research, branding has gained prominence as a crucial mechanism for marketers to distinguish themselves from their counterparts by honing their branding skills. Additionally, according to (Singh, 2004), marketers play a pivotal role in shaping a company's marketing strategy. Despite being overlooked in business plans for an extended period, branding has recently attracted the attention of marketers as a means to foster BL (Fournier and Yao, 1997). While it is highly important to create brand attachment in brand management, many companies have considered it in their strategies and are in the process of formulating strategies pursuing this goal (Bairrada and Lizanets, 2019). As consumers face a large number of brands that can expand the range of their choices, it is important for a company to be able to create and maintain BL for itself (Rizomyliotis et al., 2021; Nawaser et al., 2014; Dehkordy et al., 2013) such customers loyal to the brand are then valuable for enterprises (Chang, 2021). On the other hand, BL contributes to the steady growth of a business by increasing market share and profitability, and maintaining loyal customers is necessary for business sustainability (Ramachandran and Balasubramanian 2020). BL is also the key to success in business (Kim et al., 2021) as well as a main component of brand equity that is a competitive advantage because it seeks to develop long-term relationships with customers. BL is defined by both a customer's preference for a brand over other brands, and the customer's commitment to repurchasing the brand (Salem and Salem, 2021). According to Ramachandran (2015). BL is the proportion of each family's purchase of a brand (Tee, Gharleghi, Chan, 2013).

Examining the research conducted over the last three decades on BL structure, it can then be stated that it (BL) has two dimensions: behavioral and attitudinal loyalty. The former is related to the purchase quantity and to the number of times customers buy the brand and repeat their purchase; the latter includes some concepts such as intention to purchase and intention to recommend to others, which are both functions of customer's psychological commitment to the brand (Nam et al., 2011). In some studies, a combination of these two dimensions is considered, including both behavioral and attitudinal factors with a comprehensive view of this structure (Rather, 2018). BL dates back to more than 70 years ago, a concept that has been debated in marketing over the last few decades (Fournier and Yao, 1997; Nawaser et al., 2015; Jin and Koh 1999). BL was first introduced into the marketing literature in 1952 by Brown and became a hot topic in this respect, regarded also fanatical because it mostly dealt with branded goods (Zhou et al., 2021; Tee, et al, 2015).

By collecting knowledge in a domain and showing the current status of a discipline, literature review accordingly plays an essential role in research (Linnenluecke et al., 2020). Therefore, examining review studies in BL, the researcher found three lines of investigation in this regard: (i) a meta-analysis study by Wu and Anridho 2016, which sought to identify the effective factors and mediators of BL; (ii) a review study by Palumbo and Herbig (2000) investigated the concept of BL; and (iii) a meta-analysis study on this structure in the hotel industry (Górska-Warsewicz and Kulykovets, 2020). A meta-analysis investigation examining the relationship between BL and social network marketing activities can also be mentioned (Ibrahim, 2022). Although we can extract practical knowledge from these reviews, it should be noted that there has been no analysis depicting the future state of this structure.

Today, there is a growing knowledge production in business research, which is scattered given its huge volume (Snyder, 2019b). Survey of published works concerning BL structure (Figure 1) indicates the high number of studies in this regard, leading to knowledge accumulation; therefore, bibliography is one of the most important tools for evaluating and analyzing the output from a large number of studies in BL to appraise their quality (Cobo et al., 2015; Moezzi et al., 2012; Jahanshahi et al., 2019). Nowadays, many fields of research have reached maturity; therefore, there is growing interest to comprehensively examine different domains using advanced softwares (Koseoglu et al., 2016). In recent years, research on business strategy, finance, human resources, management, and marketing indicate the growth of bibliometric analysis, which is used to discover trends in the performance of journal authors, and in general, to determine the thinking direction in one area because of aiming at identifying the emerging trends in each domain (Donthu et al., 2021; Samadi, et al, 2015).

The current study adopted a bibliometric approach in BL to discover the trends, performance, and intellectual structure in this domain, and to draw a clear roadmap for researchers and for future research on BL. Therefore, the following goals are being sought here:

1. Examining functional components such as author, journal, country, and university in BL domain to identify valid documents and the leading countries and universities in this regard;
2. Sketching scientific roadmaps to determine the intellectual structure in BL;
3. Identifying the direction of future research in BL.

The present research is organized as follows. In the second section, the research method and its mechanism are described. In the third section, based on research objectives, results in performance, network, and content areas are described. In the fourth section, the obtained results are discussed. Finally in the fifth section, research limitations along with suggestions for future research are discussed.

## Methodology

Literature review espouses a scientific method and seeks to present a complete report in a specific area (Green, 2017). It can be observed that literature review has moved from a traditional analysis toward a systematic one over time, examines past research, and not just intending to review them, but also trying to extract appropriate answers from studies by analyzing and combining data (Márcio et al., 2016). Nowadays, concomitant with the increase of scientific productions in different domains, it is necessary to deploy an appropriate method to integrate the high volume of these scientific productions in different domains and specify the connection between them. The bibliometric approach, which has a long history in science, can use a large volume of data from past studies and classify them based on a number of indicators to draw a perspective of the relevant field for researchers. These bibliometric studies are useful to describe, evaluate, and monitor the published research (Wallin, 2005; Zupic & Čater, 2014). This feature of bibliometric differentiates it from other review studies (Fahim, Faryal, 2022) because due to the diversity of data, this approach does not have individual bias like other review studies (Gurzki & Woisetschläger, 2017). Moreover, data collection and analysis in bibliometrics allows for prediction of the future of a specific domain through discovering the hidden patterns of each area (Hashemi et al., 2022). Bibliography has become highly popular in business research nowadays, which can manage a large volume of citation data and illustrates them in the form of meaningful patterns for researchers. The softwares used in this approach are compatible with citation databases such as Scopus and WOS (web of science) to extract citation data (Mukherjee et al., 2022).

In this research, a positivist paradigm with a quantitative approach has been applied according to the systematic review of the existing literature. To follow the existing analogy in quantitative approach, the researcher pursued a specific process called research plan or strategy. In a step-by-step process, this strategy shows the path and the method to achieve the goals (Barad, 2017) so that one can reach the best results in terms of research goals from the large amount of data. The researchers thus proceed with research according to the steps of the following diagram.

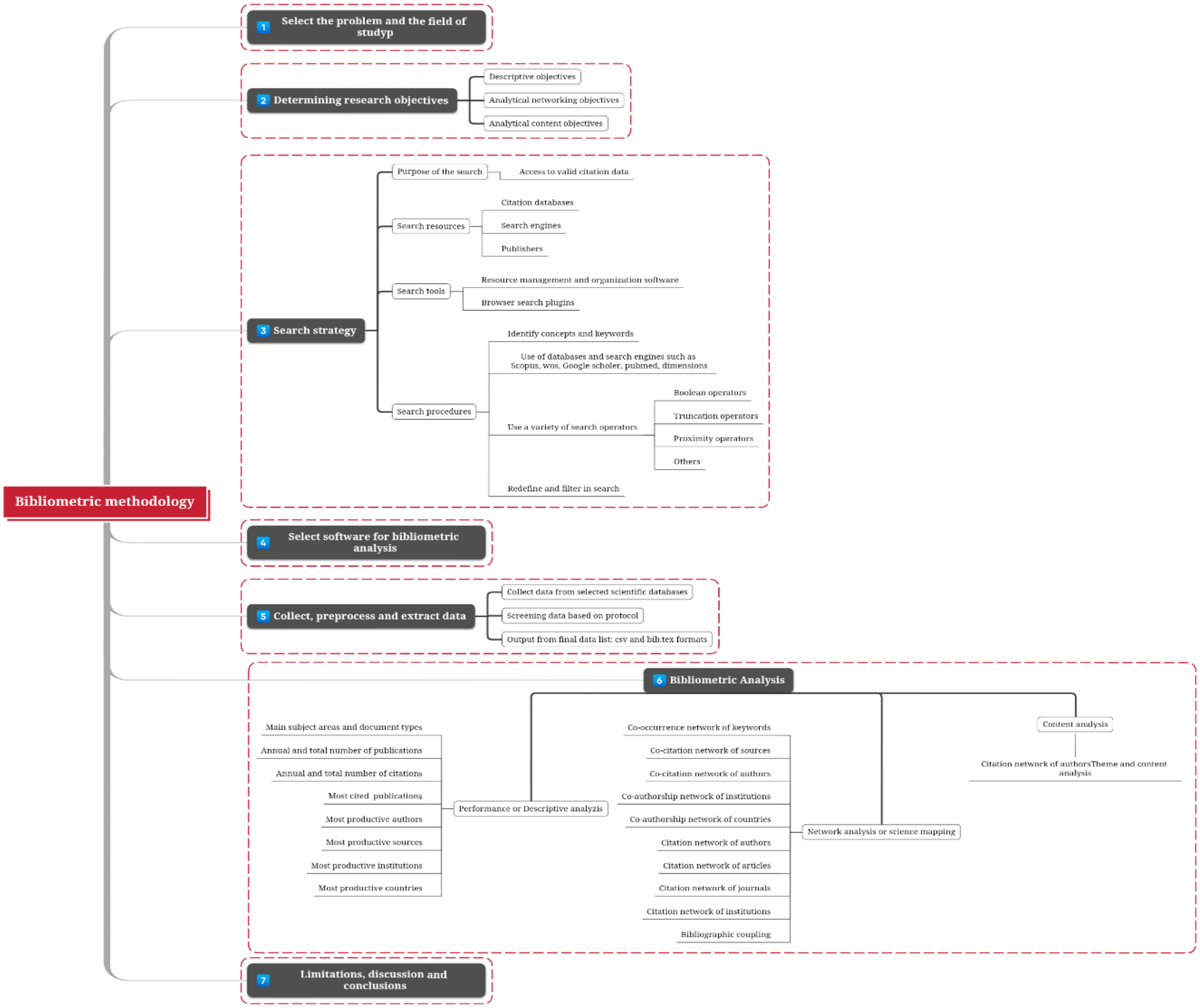

**Diagram 1**. Methodological process of bibliometric studies (Moradi and Miralmasi, 2020:570).

This study aims to review the literature on BL and its current and future state using bibliometric techniques. The authors use four steps to conduct their analysis: (i) choosing the problem and the domain of BL and its importance for firms and consumers; (ii) setting goals to examine the performance and citation network of BL research; (iii) specifying the search strategy to select 1,468 documents from Scopus database using the keyword "BL"; and (iv) choosing Publish and Perish, R (biblioshiny) and VOSviewer software to search for relevant documents, perform performance, content, co-citation, co-authorship, and co-word analysis.

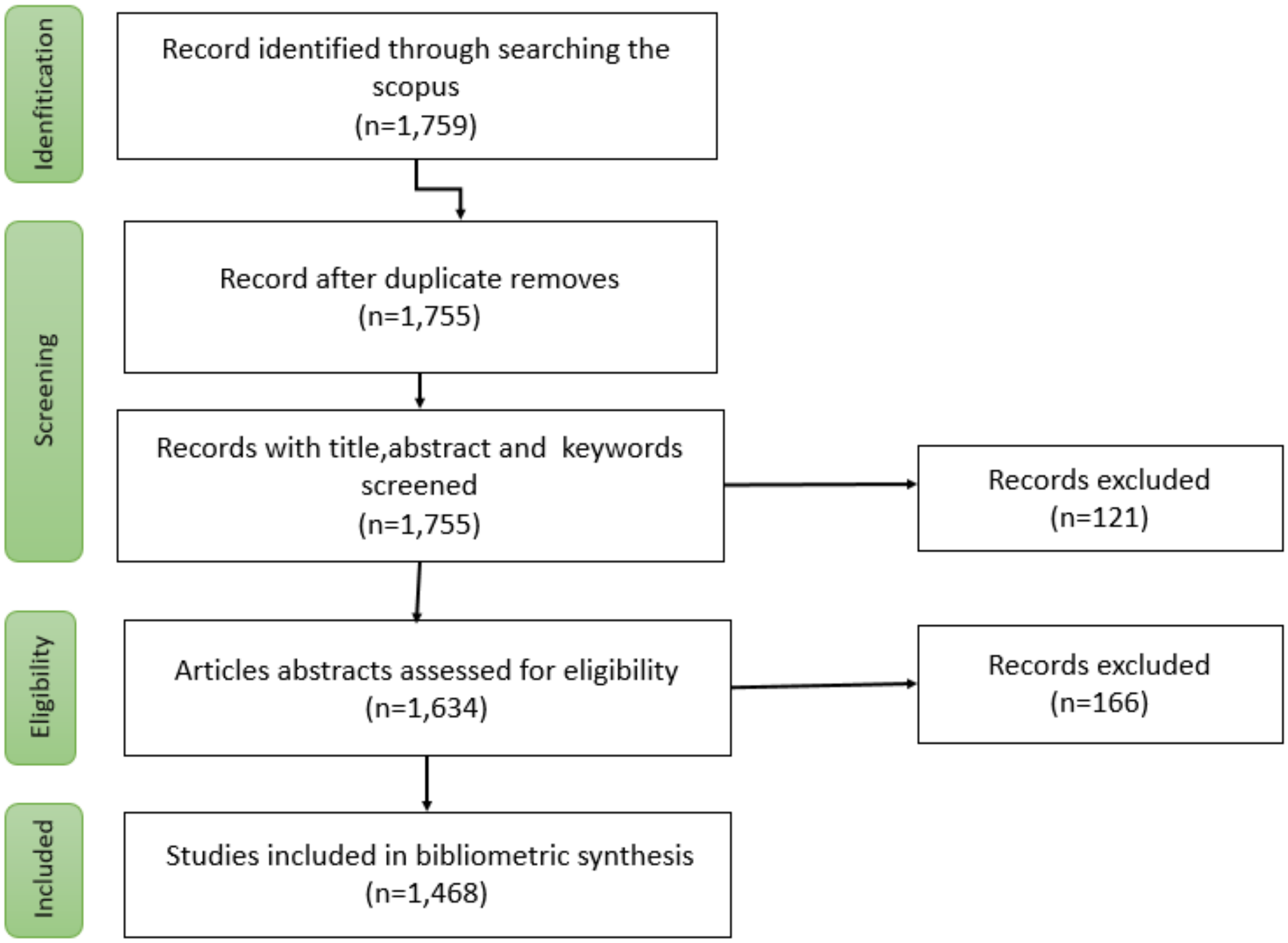

**Figure 2.** Application of the developed PRISMA protocol (Page et al., 2021)

## Results and Discussion

According to the proposed step-by-step process in diagram (1), after data extraction and mining, the researchers examined the results of bibliometric analysis in the sixth step. In general, there are two methods for bibliometric citation data analysis: performance analysis and scientific mapping. In performance analysis, a number of indices are reviewed in order to label the prevailing conditions in the respective area, which can also be considered as descriptive; in scientific mapping as a kind of network analysis, interactions and connections between citations data in that domain are investigated (Donthu et al., 2021). Finally, the researcher performs content analysis by studying the keywords in the forms of WordCloud and word growth diagram in order to introduce gaps and areas for future research.

### *Analyzing primary documents*

The citation data of research including information about authors, keywords, publications, and documents are analyzed as tabulated in Table 1, showing an overview of documents published between 1994 and 2022 in the area of BL. The articles in this domain have grown by 18.5% annually, which is appropriate for this area. Moreover, there have been on average 32.23 citations per article. Out of 1,468 articles on BL, 211 have been single-authored. There has also been collaboration with more than one author in 1257 articles with cooperation rate of 2.61, from authors of different countries in 26.29% of the articles.

**Table 1.** The main bibliographic information.

| Description | Results |
|---|---|
| Timespan | 1994-2022 |
| Sources (Journals, Books, etc.) | 376 |
| Documents | 1468 |
| Annual Growth Rate % | 18.5 |
| Document Average Age | 6.4 |
| Average citations per doc | 32.23 |
| References | 78429 |
| Keywords Plus (ID) | 630 |
| Author's Keywords (DE) | 3395 |
| Authors | 3040 |
| Authors of single-authored docs | 194 |
| Single-authored docs | 211 |
| Co-Authors per Doc | 2.63 |
| International co-authorships % | 26.29 |
| article | 1468 |

***Publication process***

Figure 1 shows the scientific production in the area of BL between years 1994 and 2022. Before 1994, the number of scientific productions was single-digit, and since 1994, we have seen an upsurge in articles in a way that scientific production increased from 10 articles in 1994 to 209 by July 2022. There have also been fluctuations in publications during these years, and the number of publications in some years, such as 2009, 2013, 2017, and 2021, decreased compared to the previous years albeit not significantly. Production in BL is growing in 2022 and in the future, more production in this area would be seen, showing research interest in BL.

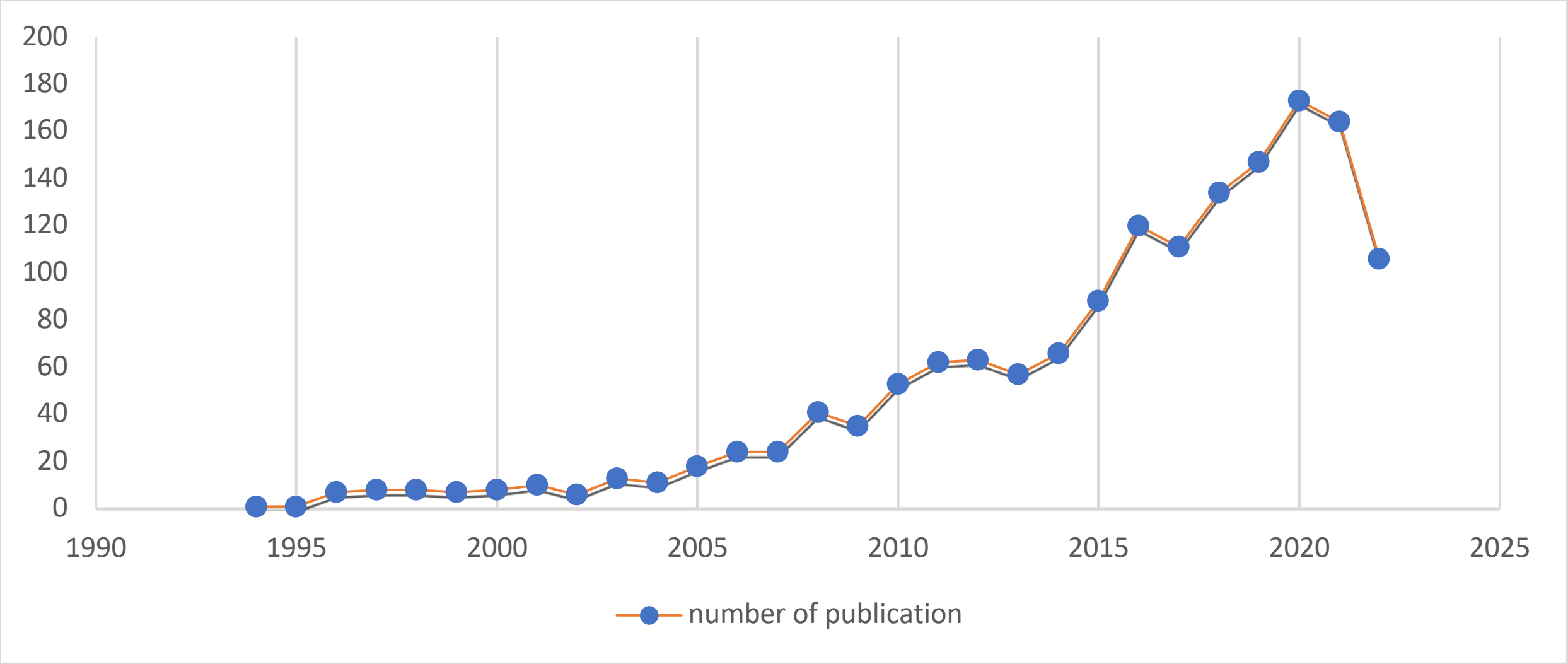

**Figure 1.** Number of BL publications by year (1994-2022).

### *Performance analysis*

Performance analysis is a basic procedure in bibliometrics, which identifies the optimal documents, authors, organizations, countries, and publications in each domain based on citation data (Cucari et al., 2022).

### *Impactful articles*

Table 2 shows 10 studies receiving the highest number of citations in BL. According to this table, the article by Batra, Ahuvia, and Bagozzi received the largest number of citations (889), cited annually by 80.82% on average since publication date. Conducted in the form of a qualitative work in three studies, this article shows that the higher-order models including the first-order model, are able to better predict brand love outcomes, namely BL. Articles by Kressmann, Sirgy, Herrmann, Huber, Huber, and Lee, and that of Pappu, Cooksey, and Quester respectively received 527 and 475 citations in the next rank. The former examines the direct and indirect effects of self-image matching on BL, while the latter investigates its dimensions with a real sample from Australia in order to better measure the brand equity with BL as one of its dimensions.

**Table 2.** Mostly cited articles on BL.

| Authors | Title | year | Journal | TC | C/Y |
|---|---|---|---|---|---|
| Batra R., Ahuvia A., Bagozzi R.P. | Brand Love | 2012 | Journal of Marketing | 889 | 80.82 |
| Kressmann F., Sirgy M.J., Herrmann A., Huber F., Huber S., Lee D.-J. | Direct and indirect effects of self-image congruence on BL | 2006 | Journal of Business Research | 527 | 31 |
| Pappu R., Cooksey | Consumer-based brand equity: | 2005 | Journal of | 475 | 26.39 |
| R.W., Quester P.G. | improving the measurement-empirical evidence | | Product & Brand Management | | |
| Sprott D., Czellar S., Spangenberg E. | The importance of a general measure of brand engagement on market behavior: Development and validation of a scale | 2009 | Journal of Marketing Research | 473 | 33.79 |
| Chen P.-Y., Hitt L.M. | Measuring switching costs and the determinants of customer retention in internet-enabled businesses: A study of the online brokerage industry | 2002 | Information Systems Research | 469 | 22.3 |
| Laroche M., Habibi M.R., Richard M.-O. | To be or not to be in social media: How BL is affected by social media? | 2013 | International Journal of Information Management | 445 | 44.5 |
| Nam J., Ekinci Y., Whyatt G. | Brand equity, BL and consumer satisfaction | 2011 | Annals of Tourism Research | 421 | 35.08 |
| Stokburger-Sauer N., Ratneshwar S., Sen S. | Drivers of consumer-brand identification | 2012 | International Journal of Research in Marketing | 410 | 37.27 |
| Yi Y., Jeon H. | Effects of loyalty programs on value perception, program loyalty, and BL | 2003 | Journal of the Academy of Marketing Science | 392 | 19.6 |
| Wang Y., po lo H., Chi R., Yang Y. | An integrated framework for customer value and customer-relationship-management performance: A customer-based perspective from China | 2004 | Managing Service Quality: An International Journal | 353 | 18.58 |

### *Impactful authors and journals*

This study also identifies the most influential authors and journals in BL research, based on their productivity and impact. Productivity is measured by the number of articles published by an author, and impact is quantified by the number of citations received by an article (Filho et al., 2022). Figure 3 shows the top 10 authors in BL research, along with their number of articles and citations. Bagozzi is the most cited author with 983 citations from two articles, followed by Wang with 947 citations from five articles.

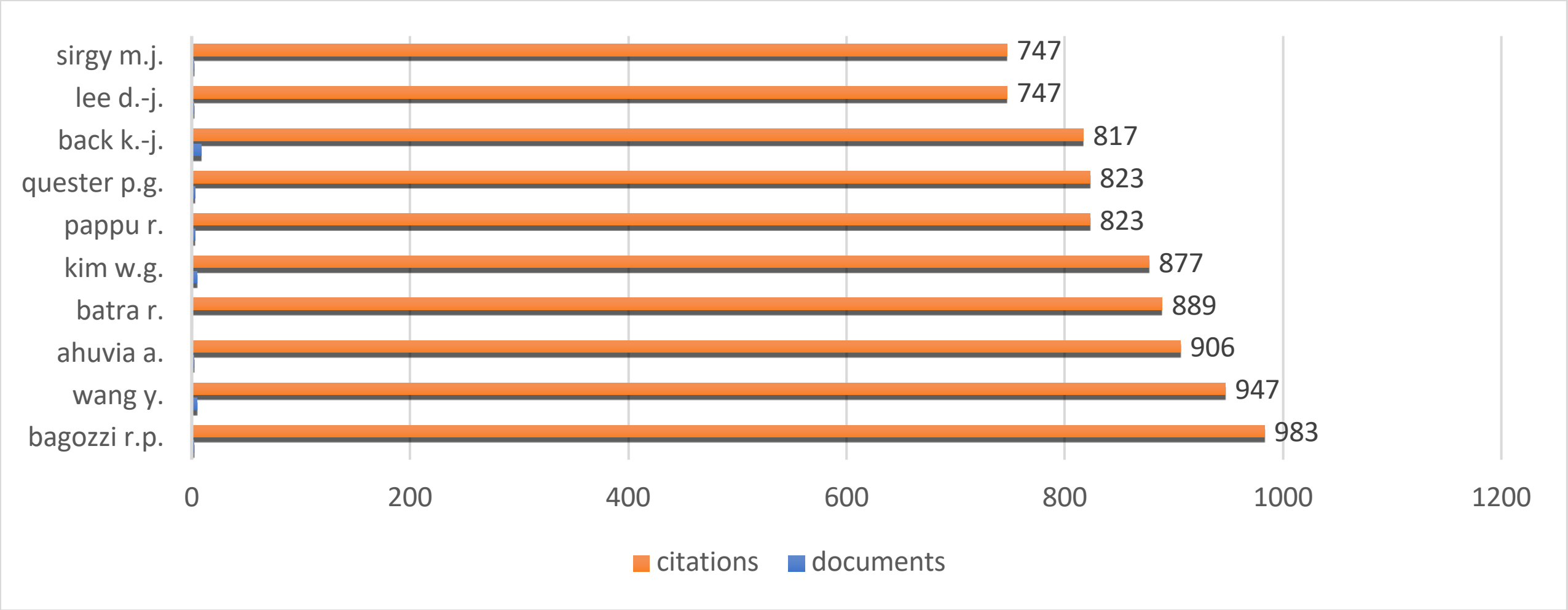

**Fig 3.** Top-10 most influential authors in BL

**Table 3:** Metrics on top-10 production sources

| Sources | h_index | g_index | m_index | TC | NP | PY_start |
|---|---|---|---|---|---|---|
| JOURNAL OF BUSINESS RESEARCH | 33 | 61 | 1.5 | 4147 | 61 | 2001 |
| JOURNAL OF PRODUCT AND BRAND MANAGEMENT | 32 | 59 | 1.778 | 3528 | 64 | 2005 |
| EUROPEAN JOURNAL OF MARKETING | 19 | 34 | 0.76 | 1733 | 34 | 1998 |
| JOURNAL OF PRODUCT & BRAND MANAGEMENT | 14 | 17 | 0.56 | 1678 | 17 | 1998 |
| JOURNAL OF CONSUMER MARKETING | 18 | 28 | 0.783 | 1624 | 28 | 2000 |
| JOURNAL OF BRAND MANAGEMENT | 18 | 33 | 1.385 | 1540 | 33 | 2010 |
| JOURNAL OF RETAILING AND CONSUMER SERVICES | 22 | 39 | 0.815 | 1537 | 41 | 1996 |
| INTERNATIONAL JOURNAL OF HOSPITALITY MANAGEMENT | 14 | 20 | 1.077 | 1039 | 20 | 2010 |
| JOURNAL OF FASHION MARKETING AND MANAGEMENT | 14 | 19 | 0.583 | 656 | 19 | 1999 |
| ASIA PACIFIC JOURNAL OF MARKETING AND LOGISTICS | 14 | 21 | 0.56 | 462 | 25 | 1998 |

Table 3 shows 10 publications in BL based on the number of their citations and indices such as h-index, g-index, and m-index. According to the number of citations, Journal of Business Research (4147), Journal of Product and Brand Management (3528), European Journal of Marketing (1733), Journal of Product and Brand Management (1678), Journal of Consumer Marketing (1624) are the top five publications in this field. The h-index evaluating scientific resources based on their productivity (Boakye et al., 2022) shows that Journal of Business Research has the highest number of citation (g-index=33), followed by Journal of Product and Brand Management with g-index=32; and Journal of Retailing and Consumer Services with g-index=22 and 1537 citations.

***Top organizations***

Table 4 shows 10 organizations that had the most and the least scientific productions in BL out of 1341 organizations. According to this table, University of South Australia is the leader in this regard with the production of 34 articles, and only two universities, namely University of Valencia from Spain and University of Coimbra from Portugal, are among the 10 institutions with the largest number of productions; the other seven institutions are from Australia and South Korea, indicating that BL is favored by Asian researchers. 724 institutions, out of 1341, namely 53% of all institutions working in this domain, had only one scientific production in this field, 10 of which are tabulated in Table 4.

**Table 4.** Ten organizations having the most and the least scientific productions in the field of BL

| **Most Article** | | **Lowest Article** | |
|---|---|---|---|
| **Affiliation** | **Articles** | **Affiliation** | **Articles** |
| UNIVERSITY OF SOUTH AUSTRALIA | 34 | WORLD ISLAMIC SCIENCES AND EDUCATION UNIVERSITY (WISE) | 1 |
| SEJONG UNIVERSITY | 30 | XI'AN JIAOTONG UNIVERSITY | 1 |
| ISLAMIC AZAD UNIVERSITY | 25 | XIÂ€™AN JIAOTONG UNIVERSTIY | 1 |
| UNIVERSITY OF VALENCIA | 23 | XIAMEN UNIVERSITY MALAYSIA | 1 |
| SEOUL NATIONAL UNIVERSITY | 22 | XLRI SCHOOL OF BUSINESS AND HUMAN RESOURCES | 1 |
| SWINBURNE UNIVERSITY OF TECHNOLOGY | 21 | YAZD BRANCH OF PAYAM NOOR UNIVERSITY | 1 |
| HANYANG UNIVERSITY | 19 | YORK UNIVERSITY | 1 |
| CURTIN UNIVERSITY | 17 | YOUNGSTOWN STATE UNIVERSITY | 1 |
| UNIVERSITY OF COIMBRA | 17 | YU DA UNIVERSITY | 1 |
| UNIVERSITY OF SOUTH CAROLINA | 16 | ZAGAZIG UNIVERSITY | 1 |

***The most effective countries***

Table 5 shows the countries having the lowest and highest productions in BL. The United States has the highest scientific production in this domain with 834 articles. The composition of countries in terms of scientific production in BL shows that Asian countries (India, China, South Korea, Indonesia, and Iran) have more productions in BL than European countries (England and Spain), indicating that Asian researchers are more interested in this area as shown in Table 4.

Table 6 shows the lowest and highest citations associated to these countries. The United States ranked first with 11131 citations followed by South Korea (4019), England (3064), Australia (2929) and Spain (2387). We also observed that small and less developed countries (except Japan) have received the least citations in this field.

**Table 5.** Ten countries with the lowest and highest production in BL

| Most Article | | lowest Article | |
|---|---|---|---|
| **country** | **Article** | **Country** | **Article** |
| USA | 834 | ZAMBIA | 2 |
| INDIA | 388 | ALBANIA | 1 |
| CHINA | 306 | BAHRAIN | 1 |
| UK | 263 | BULGARIA | 1 |
| AUSTRALIA | 262 | GEORGIA | 1 |
| SOUTH KOREA | 214 | ICELAND | 1 |
| MALAYSIA | 173 | KENYA | 1 |
| SPAIN | 158 | MALTA | 1 |
| INDONESIA | 94 | PANAMA | 1 |
| IRAN | 90 | YEMEN | 1 |

**Table 6:** Ten countries with the most and least citations in BL

| Most TC | | | lowest TC | | |
|---|---|---|---|---|---|
| **Country** | **TC** | **Average Article Citations** | **Country** | **TC** | **Average Article Citations** |
| USA | 11131 | 48.19 | PHILIPPINES | 4 | 4.00 |
| KOREA | 4019 | 50.87 | SRI LANKA | 3 | 3.00 |
| UNITED KINGDOM | 3064 | 49.42 | JAPAN | 2 | 0.50 |
| AUSTRALIA | 2929 | 39.05 | ROMANIA | 2 | 1.00 |
| SPAIN | 2387 | 51.89 | TRINIDAD AND TOBAGO | 2 | 2.00 |
| CHINA | 2256 | 25.64 | ZAMBIA | 1 | 1.00 |
| INDIA | 1241 | 14.60 | ZIMBABWE | 1 | 0.33 |
| CANADA | 1042 | 61.29 | BOSNIA | 0 | 0.00 |
| FRANCE | 881 | 51.82 | ISRAEL | 0 | 0.00 |
| MALAYSIA | 806 | 20.67 | MOROCCO | 0 | 0.00 |

### *Network analysis and scientific maps*

Cluster network maps show the development of a field over time and provide an objective description of different aspects achieved by researchers in various periods of time. Increasing access to Internet databases containing citation data and developing better analytical tools could also enhance their appeal among academics and researchers (Kumar et al., 2022).

### *Co-word analysis*

In co-word analysis, unlike other analyses of bibliometric networks focusing on citation, the word itself is analyzed. Through this analysis, which mostly concerns author's keywords, the actual content of documents is accordingly checked, and it is assumed that the words that come together in most documents are thematically related.

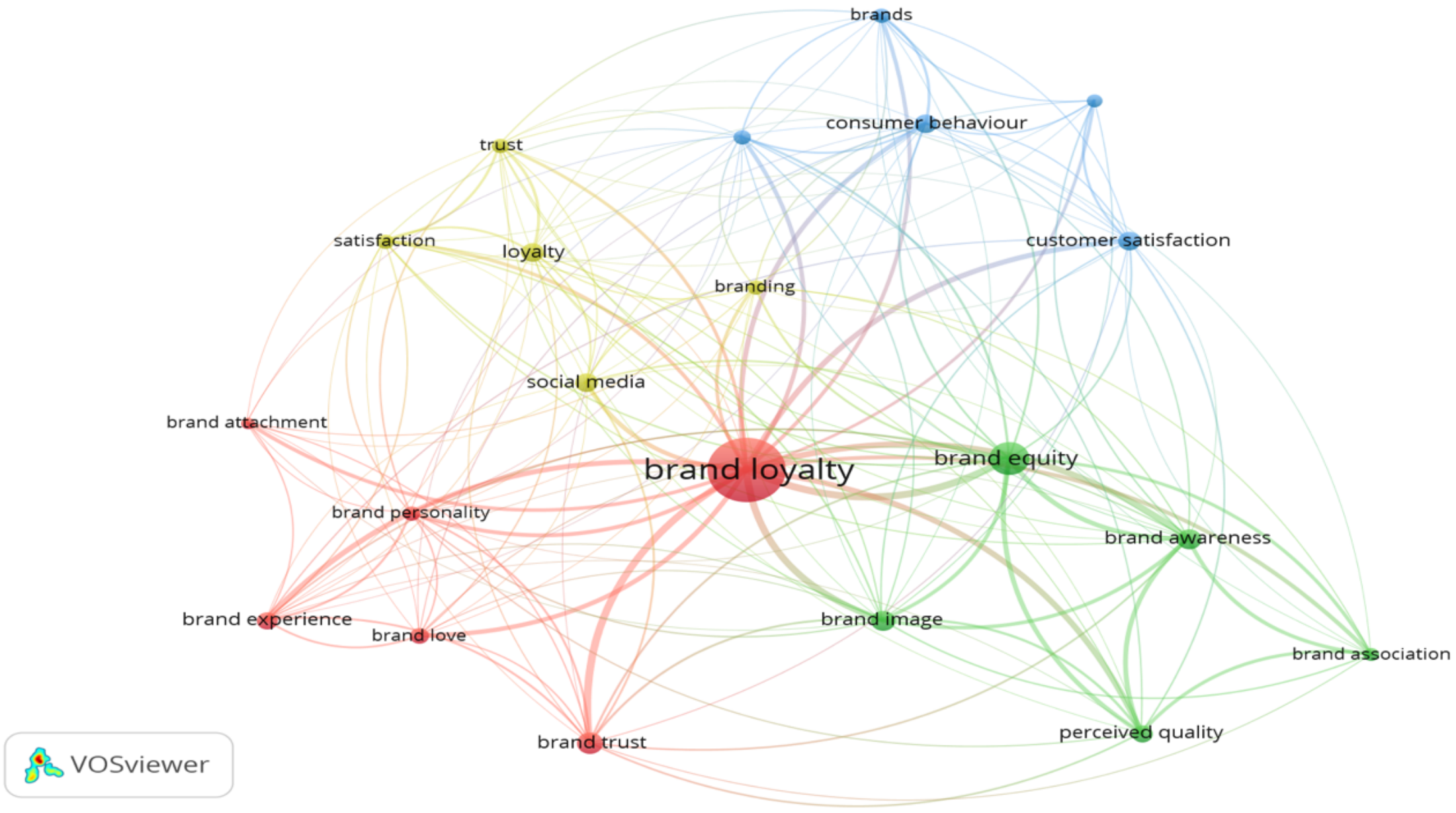

**Fig. 4.** Map of keywords with co-word analysis

Figure 4 shows the network of keywords with at least 30 co-occurrences with "BL". Each node represents a keyword, and its size reflects its frequency (Donthu et al., 2021). The main keyword, "BL", is the largest node with 784 occurrences, followed by "brand equity" with 220 occurrences. Each link represents the co-occurrence of two keywords, and its thickness reflects the number of co-occurrences. For example, "BL" and "brand equity" have a thick link because they are frequently used together. Each color represents a cluster of related keywords. There are four clusters in red, green, blue, and yellow. The red cluster includes keywords related to the emotional aspects of BL, such as "brand love", "brand attachment", and "brand trust". The green cluster includes keywords related to the cognitive aspects of BL, such as "brand association", "brand image", and "perceived quality". The blue cluster includes keywords related to the behavioral aspects of BL, such as "brand management", "consumer behavior", and "customer loyalty". The yellow cluster includes keywords related to the social aspects of BL, such as "branding", "social media", and "trust".

### *Co-authorship analysis*

Analysis of co-authorship is common in bibliometrics, indicating the degree of interactions between researchers, organizations, and countries in a domain, as well as the kinds of network created by these interactions for cooperation (Aria & Cuccurullo, 2017). According to the number of citations and articles shared by authors, VOSviewer software draws a network of authors, showing which authors have published the most collaborative scientific documents in a particular field (Patel et al., 2022).

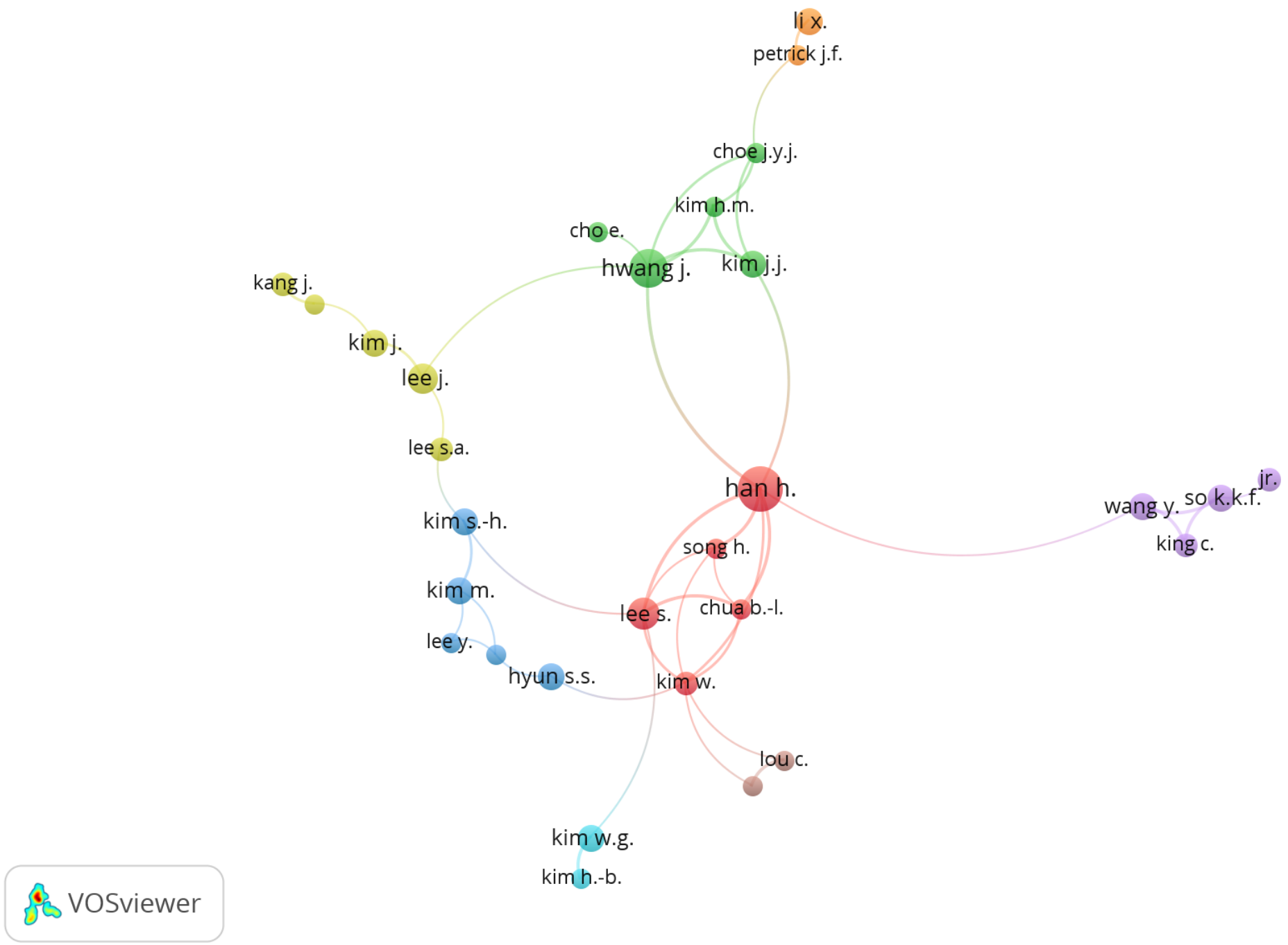

**Fig 5.** Author collaboration network in BL

Figure 5 shows a network of collaborations between authors based on possessing at least three documents in this field provided that their documents have at least one citation. A total of 168 authors were estimated for participation. In this figure, each node represents an author, and the size of nodes is determined by the number of articles published by the author. Eight clusters have accordingly been created showing that there are collaborations between eight research teams in these domains, and that the red cluster and its subsets (Chuba, Kim W, Lee S, Song H) have the most significant collaboration followed by Hwang's green cluster and its subsets (Choe, Kim, and Ki). This figure also shows that the collaborations are scattered and perhaps we see only two authors in these clusters. Due to large distances geographically, the authors seem not to work closely together in this field.

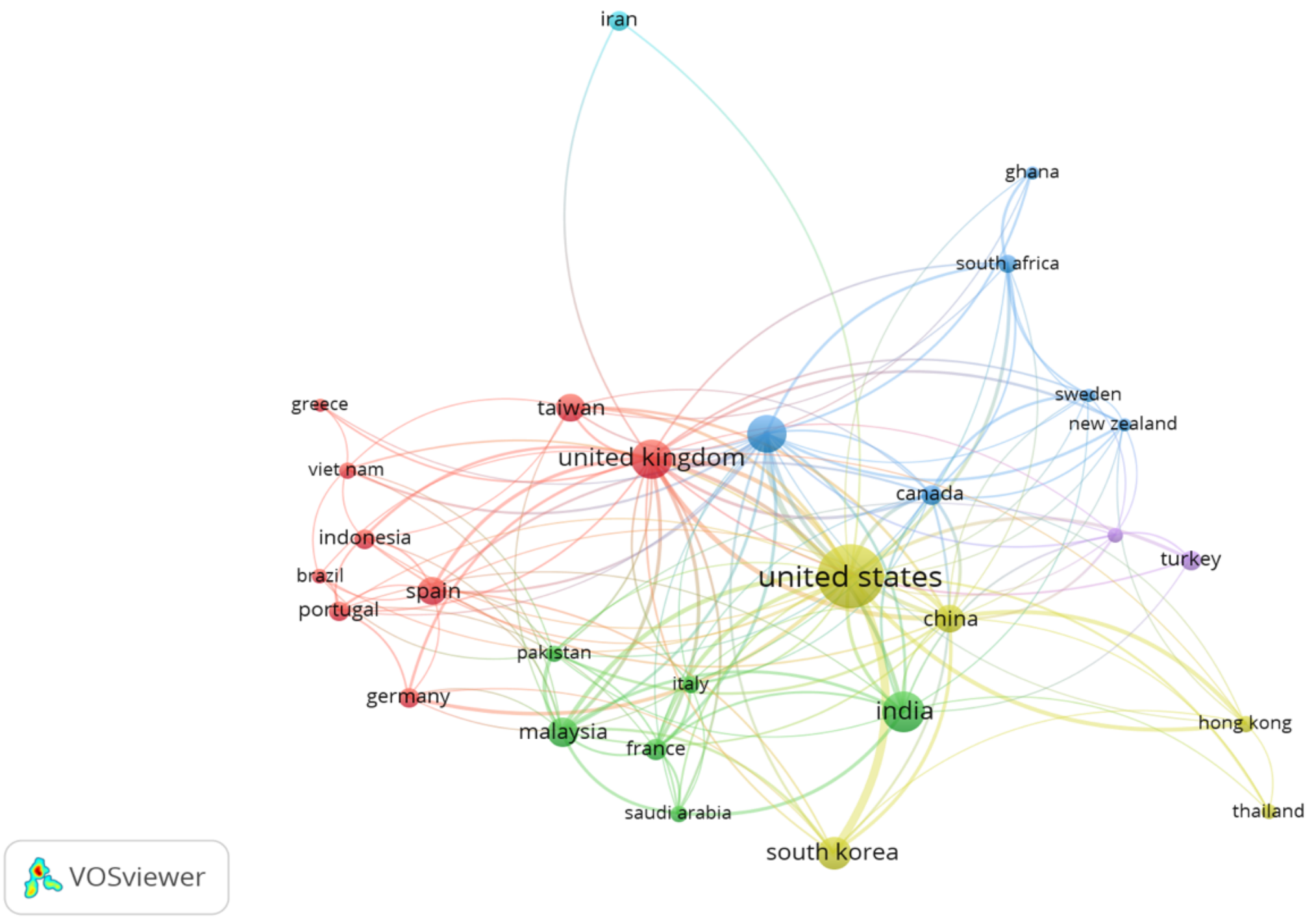

**Fig. 6.** Cooperation network of authors from countries producing articles in BL

Figure 6 shows the collaboration network of countries that produce articles on BL. The map includes 29 countries that have co-authored at least 15 cited articles on BL. Malaysia and Taiwan are the most active in international collaboration, while the USA and the UK have partnered with all the countries on the map. The USA and South Korea have the highest level of collaboration and belong to the same cluster, colored in yellow. The USA also has strong ties with Australia, Germany, Malaysia, and China. The collaboration between countries in this research area does not depend on their language similarity or geographical proximity.

### *Co-citation analysis*

Figure 7 shows the co-citation network of authors who have received at least 100 citations on BL, based on 181 prominent authors in BL. The network reveals the hidden patterns and relationships between authors in this domain (Patel et al., 2022). Each node represents an author, and its size reflects the number of citations. Each link represents the co-citation of two authors, and its thickness reflects the frequency of co-citation. Each color represents a cluster of related authors. The most co-cited authors are Keller, Hair, Aaker, and Fornel, in descending order. The map consists of four clusters in blue, red, green, and yellow.

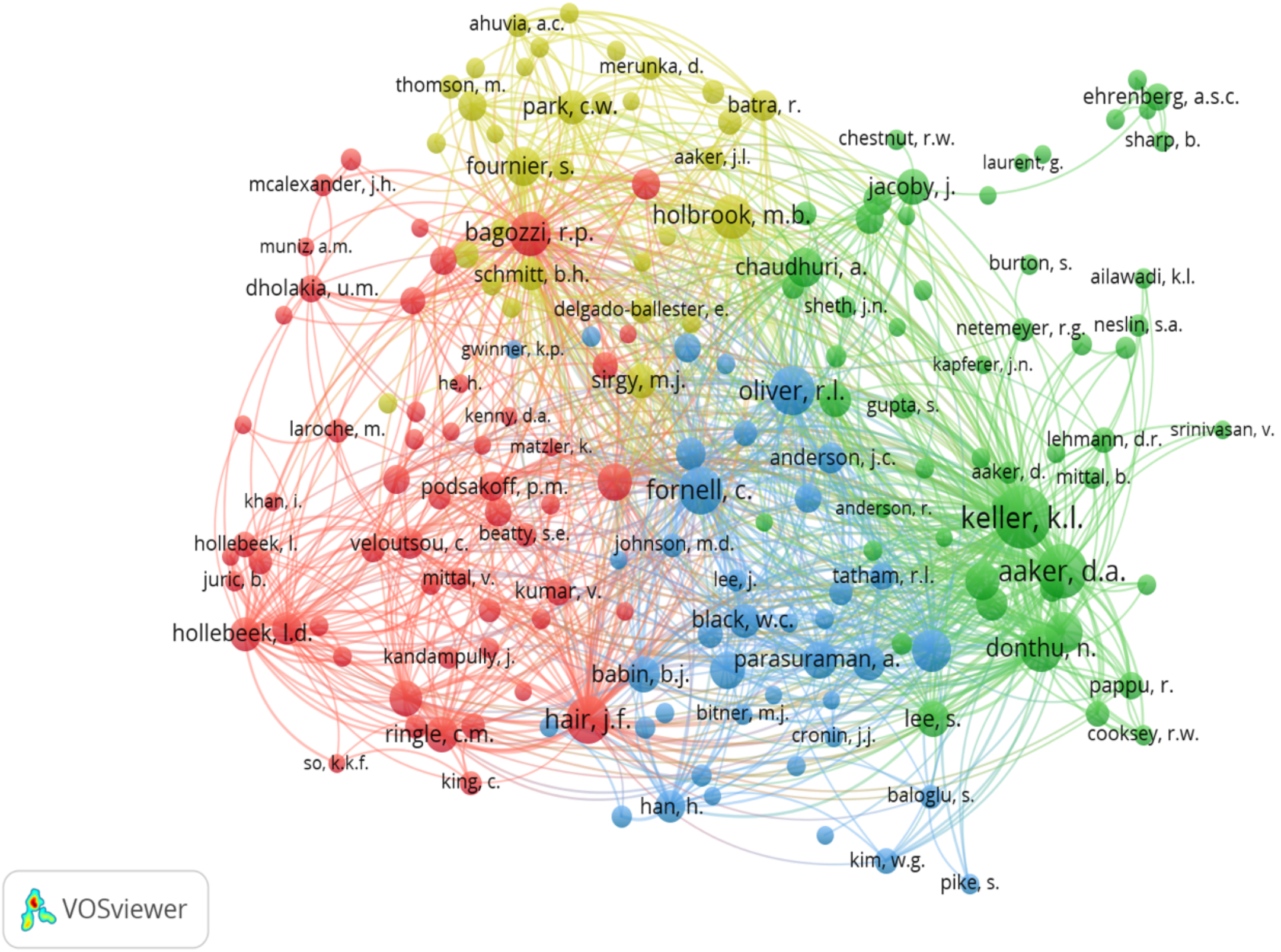

**Fig. 7.** Co-citation network of BL authors

### *Identification of trends and orientations of future research in BL*

This study analyzes the keywords in abstracts, titles, and keywords of BL articles to identify the research gap and the future trends in this field (Kashani et al., 2022). The keywords are classified into author and index keywords, which reflect the content and the subject of the articles, respectively (Kumar & Harichandan, 2022). The growth of author keywords over 15 years reveals the roadmap and orientation of BL research. The highly cited articles in the last four years with more than 20 citations are also evaluated to examine their recent trends and contents.

### 3.5.1 WordCloud

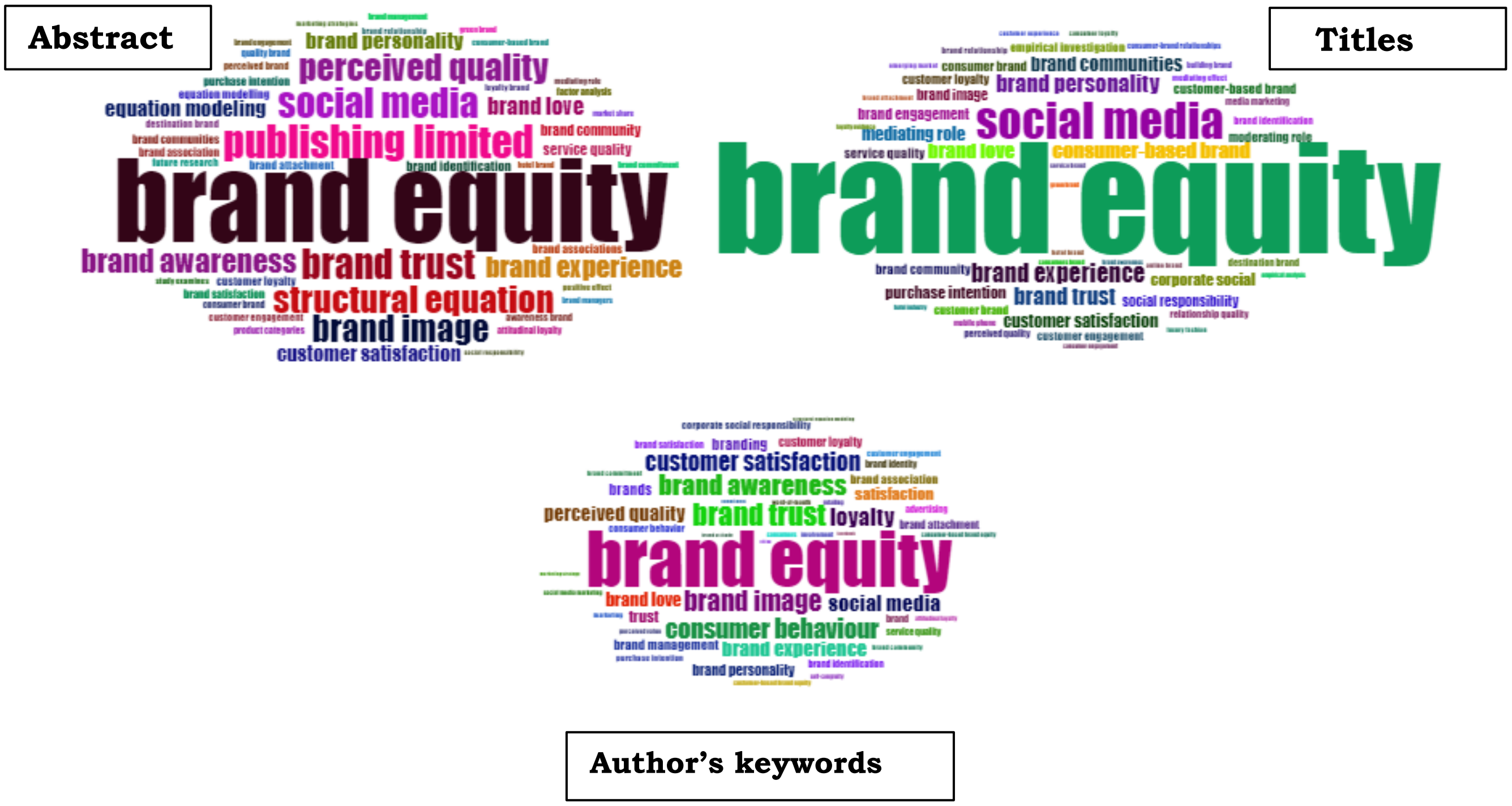

**Fig. 8.** Words with the highest frequency in abstract, title, and keyword.

Figure 8 shows the WordCloud of the 50 most frequent words in abstracts, titles, and keywords of BL articles. The word size reflects the frequency of occurrence. The most frequent words are loyalty, brand equity, publishing limited, brand trust, and social media. Other words are shown in the figure in ascending order of frequency. The analysis suggests that future research should focus on the less frequent words, such as Facebook, commitment, and marketing strategy, as they represent relatively understudied areas.

***Thematic evolution of keywords or word growth***

Figure 9 shows the word growth of author keywords related to BL from 1994 to 2022. The keywords with sudden surge indicate the research interest and trend in that domain (Rejeb et al., 2022). The figure shows that "brand experience" and "social media" have been popular topics since 2011, while other keywords, such as "brand equity", "brand trust", "brand image", "brand awareness", and "consumer behavior" have been steadily growing. The figure also shows that COVID-19 has not been considered as a keyword in BL research, which suggests a potential gap for future studies.

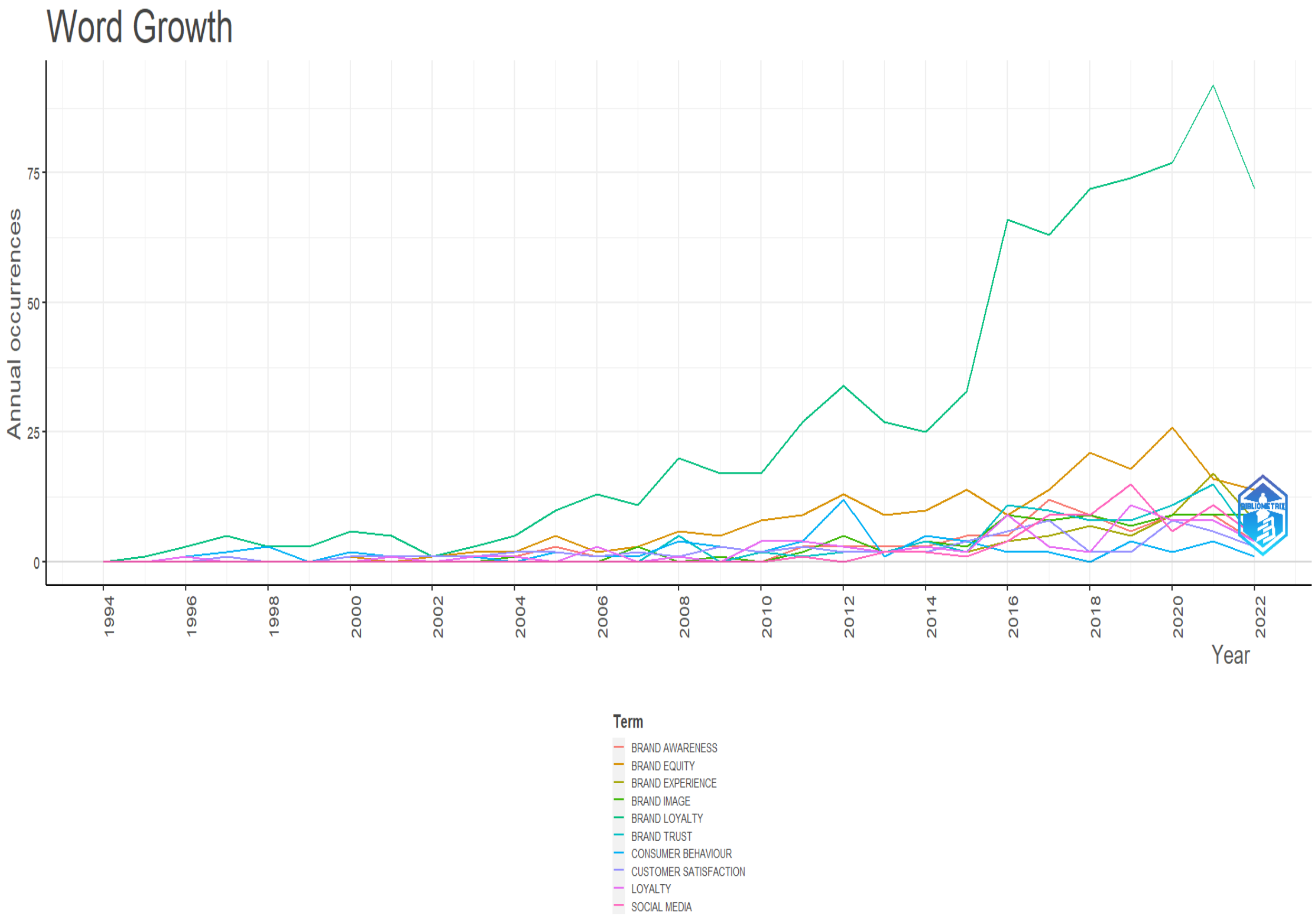

**Fig 9.** The evolution dynamics of BL authors keywords over time (1994-2022)

## Content analysis

This study analyzes the content of BL articles published in the last four years (2019-2022) that received more than 20 citations per year. The articles were mostly empirical and quantitative, and used SEM for data analysis. The most common sample was customers and consumers, followed by tourists, employees, social media members, and students. The most common antecedents of BL were brand trust, satisfaction, brand awareness, and perceived quality. The most common outcomes of BL were green brand evangelism, overall brand equity, purchase intention, green purchase behavior, overall destination brand equity, and WOM. The majority of the articles addressed tourism and hoteling industry. Table 7 provides further details on these articles.

**Table 7.** Top cited papers in the area of BL (2019-2022).

| Ref | TC | Sample | Research method | Research design | Data analysis approach | Consequence | Antecedent |
|---|---|---|---|---|---|---|---|
| (Spieth, Roeth, and Meissner 2019) | 21 | Customer N=510 | Empirical | Quantitative | SEM | | Value Offering Innovation-Brand Trust |
| (So, Wei, and Martin 2021) | 23 | Customer N=556 | Empirical | Quantitative | ANOVA | | |
| (Panda et al. 2020) | 82 | Buyer N=340 | Empirical | Quantitative | SEM(M-plus) | Green brand evangelism | Altruism- Attitudinal green purchasing intention |
| (Muniz et al. 2019) | 29 | Students N=409 | Empirical | Quantitative | t-test | Overall Brand Equity | CSR communication |
| (Liu and Jiang 2020) | 21 | staff in star rated hotels in China n=622 | Empirical | Quantitative | SEM(Amos) | | Brand Awareness-Brand Image |
| (Lin, Lobo, and Leckie 2019) | 25 | Student N=826 | Empirical | Quantitative | SEM(M-plus) | | Green brand innovativeness-Green perceived value |
| (Li, Teng, and Chen 2020) | 50 | Customer N=298 | Empirical | Quantitative | SEM(Amos) | | customer trust-customer engagement-brand attachment |
| (Kim, Lee, and Lee 2020) | 20 | Citizens N=1001 | | | | | |
| (Kim and Lee 2019) | 52 | members of social media n=252 | Empirical | Quantitative | SEM(Amos) | Purchase Intention | Attitude |
| (Khan, Hollebeek, Fatma, Islam, and Rahman 2020) | 46 | users of virtual service brands(consumers) n=414 | Empirical | Quantitative | SEM(Amos) | | Brand engagement via brand commitment-Brand engagement via brand trust/ brand commitment- Brand experience via brand commitment- Brand experience via brand trust/brand commitment- Brand trust via brand |

| Ref | TC | Sample | Research method | Research design | Data analysis approach | Consequence | Antecedent |
|---|---|---|---|---|---|---|---|
| | | | | | | | commitment |
| (Jung, Kim, and Kim 2020) | 21 | Consumer N=272 | Empirical | Quantitative | SEM(Amos) | | Brand image-Customer Satisfaction-Customer trust |
| (Jun and Yi 2020) | 28 | social media users n=282 | Empirical | Quantitative | SEM(Amos) | | Followers' emotional attachment-Brand trust |
| (Japutra and Molinillo 2019) | 52 | Consumer N=347 | Empirical | Quantitative | SEM (smart PLS) | | Responsible Active |
| (Fang 2019) | 51 | app users n=634 | Empirical | Quantitative | SEM (smart PLS) | | Brand Restaurant Brand Satisfaction - Value-in-use |
| (Dirsehan and Cankat 2021) | 20 | MFO ( mobile food ordering apps) n=217 | Empirical | Quantitative | SEM (smart PLS) | | MFOA Satisfaction-Restaurant Brand Satisfaction |
| (Dessart, Aldás-Manzano, and Veloutsou 2019) | 39 | members of Facebook n=970 | Empirical | Quantitative | SEM (smart PLS) | | Affective brand engagement-Behavioral brand engagement-Cognitive brand engagement! |
| (Chen et al. 2020) | 41 | Consumer N=261 | Empirical | Quantitative | SEM(Amos) | Green purchase behavior | Greenwash |
| (Bairrada, Coelho, and Lizanets 2019) | 40 | clothing brand consumers n=478 | Empirical | Quantitative | SEM(Amos) | | Brand love-Brand personality |
| (Atulkar 2020) | 20 | mall shoppers n=332 | Empirical | Quantitative | SEM (smart PLS) | | Perceived quality-Perceived Value-Perceived Satisfaction-Perceived Differentiation-Brand trust |
| (Algharabat et al. 2020) | 97 | consumers who are fans of the social media pages n=500 | Empirical | Quantitative | SEM(Amos) | | Cognitive processing-Affection-Activation Processing-Consumer Participation-Brand awareness/associations-positively-Consumer involvement-Consumer participation |
| (Abou-Shouk and Soliman | 24 | Manager N=312 | Empirical | Quantitative | SEM(PLS) | | Brand awareness-customer engagement |

| Ref | TC | Sample | Research method | Research design | Data analysis approach | Consequence | Antecedent |
|---|---|---|---|---|---|---|---|
| 2021) | | | | | | | |
| (Hwang and J. H. (Jay) Lee 2019) | 86 | senior tourists n=323 | Empirical | Quantitative | SEM(Amos) | | Consumer attitudes toward a brand-Well-being perception-Brand attachment |
| (Amoako et al. 2022) | 11 | Customer N=626 | Empirical | Quantitative | SEM(PLS) | | BL as a moderator in relationships between green marketing and purchase intentions, green marketing and sustainability with price, advertising and sustainability with price, advertising and purchase intentions |
| (Shanahan, Tran, and Taylor 2019) | 77 | User Facebook (Amazon N= Mechanical Turk.) N=242 | Empirical | Quantitative | SEM(PLS) | | Brand attachment-Consumer brand engagement--Perceived quality |
| (Song, Wang, and Han 2019) | 59 | Consumer N=401 | Empirical | Quantitative | SEM(Amos) | | Satisfaction-brand trust |
| (Khan and Fatma 2019) | 28 | Consumer: interviews N=18 + N=354 | Empirical | qualitative + Quantitative | direct interview method of qualitative research + SEM(Amos) | | Corporate social Responsibility-Brand Experience-Brand trust |
| (Mody and Hanks 2020) | 49 | Customer N=1,256 | Empirical | Quantitative | SEM(Amos) | | Brand love |
| (Tran et al. 2019) | 27 | domestic tourists n=319 | Empirical | Quantitative | SEM(Amos) | Overall destination brand equity | Destination brand image-destination perceived quality |
| (Coelho, Bairrada, and Peres 2019) | 64 | Customer N=510 | Empirical | Quantitative | SEM(Amos) | WOM-brand advocacy | brand community (identification- commitment) |
| (Liu et al. 2020) | 33 | N=1,603 | Empirical | Quantitative | ANOVA | | Affective place image-place attachment |
| (Koo, Yu, and Han 2020) | 38 | Customer N=333 | Empirical | Quantitative | SEM(Amos) | | Switching cost related to a loyalty program-Affective Commitment- |

| Ref | TC | Sample | Research method | Research design | Data analysis approach | Consequence | Antecedent |
|---|---|---|---|---|---|---|---|
| | | | | | | | Lack of the attractiveness of alternatives; |
| (Zollo et al. 2020) | 62 | Student N=420 | Empirical | Quantitative | SEM(PLS) | | Social Media Marketing Activities-Brand Experience-Social Media Benefits |
| (Hasan, Shams, and Rahman 2021) | 20 | Consumer N=675 | Empirical | Quantitative | SEM(PLS) | | Trust-interaction-perceived Risk-Novelty Value |
| (Dedeoğlu et al. 2019) | 33 | domestic and foreign tourists n=478 | Empirical | Quantitative | SEM(Amos) | | Destination brand satisfaction; |
| (Abbes, Hallem, and Taga 2020) | 25 | Consumer N=214 | Empirical | Quantitative | SEM(Amos) | | platform loyalty intentions-satisfaction-ease of use-perceived usefulness-entertainment-community belonging-seller reputation-third-party recognition |
| (Kumar and Kumar 2020) | 45 | Members N=925 | Empirical | Quantitative | SEM(Amos) | | Brand community engagement-brand community commitment |
| (Akram et al. 2020) | 30 | mobile netizens n=936 | Empirical | Quantitative | SEM(Amos) | Platform ease of using | |
| (I Khan et al. 2020) | 46 | Customer N=423 | Empirical | Quantitative | SEM(Amos) | | Customer brand experience- Calculative commitment- Customers' affective brand commitment |
| (Mody, Hanks, and Dogru 2019) | 57 | Customer N=1,256 | Empirical | Quantitative | SEM(Amos) | | Brand love-memorability |
| (Moise et al. 2019) | 29 | Guest N=378 | Empirical | Quantitative | SEM(PLS) | satisfaction | Green practices |
| (Helme-Guizon and Magnoni 2019) | 27 | Customer N= | Empirical | Quantitative | SEM(PLS) | | social brand engagement (SBE) consumer brand engagement (CBE) |
| (Schivinski et al. 2021) | 28 | consumer n=489 | Empirical | Quantitative | SEM (MPLUS | Consumption-Con- | Brand associations-Perceived quality |

| Ref | TC | Sample | Research method | Research design | Data analysis approach | Consequence | Antecedent |
|---|---|---|---|---|---|---|---|
| | | | | | | tribution | |
| (Oliveira and Fernandes 2022) | 10 | followers on Instagram N=243 | Empirical | Quantitative | SEM(PLS) | | consumer brand engagement (CBE |
| (Rather, Hollebeek, and Islam 2019) | 115 | Customer N=310 | Empirical | Quantitative | SEM(AMOS) | | Customer engagement |
| (Hwang and J. H. Lee 2019) | 23 | Senior tourists N=331 | Empirical | Quantitative | SEM(AMOS) | | Brand attachment-brand prestige of package tour-well-being perception |
| (Martínez and Nishiyama 2019) | 36 | Customer N=382 | Empirical | Quantitative | SEM(AMOS) | | CSR-Brand Awareness-Brand Image-Perceived quality |
| (Ebrahim 2020) | 52 | User N=287 | Empirical | Quantitative | SEM(AMOS) | | Brand trust-brand equity-social media marketing activities (SMM) |
| (Alnawas and Hemsley-Brown 2019) | 34 | Customer N=420 | Empirical | Quantitative | SEM(AMOS) | | Customer Experience Quality |
| (Bianchi, Bruno, and Sarabia-Sanchez 2019) | 54 | Customer N=429 | Empirical | Quantitative | SEM(AMOS) | purchase intention. - corporate reputation. | Cognitive satisfaction-Affective satisfaction |
| (Šerić and Gil-Saura 2019) | 21 | Hotel gust N=360 | Empirical | Quantitative | SEM(PLS) | Brand equity | |
| (Busser and Shulga 2019) | 29 | Customer N=492 | Empirical | Quantitative | SEM(AMOS) | Trust | Transparency- Authenticity- Involvement with CGA |
| (Luo et al. 2019) | 55 | Customer N=765 | Empirical | Quantitative | SEM(LISREL) | | Affective response-Service quality- |
| (Savastano et al. 2019) | 44 | store managers and employees N=80 | Empirical | Qualitative | QSR Nvivo (content analysis) | | Customer Satisfaction - Increased Number of Store Visits |
| (Kumar and Nayak 2019) | 39 | brand community members | Empirical | Quantitative | SEM(Amos) | | Brand attachment- customer brand engagement |

| Ref | TC | Sample | Research method | Research design | Data analysis approach | Consequence | Antecedent |
|---|---|---|---|---|---|---|---|
| | | n=282 | | | | | |
| (Fernandes and Moreira 2019) | 60 | Consumer N=655 | Empirical | Quantitative | SEM(Amos) | | consumer brand engagement-satisfaction |
| (Jin, Yoon, and Lee 2019) | 20 | Customer N=781 | Empirical | Quantitative | SEM(EQS) | | Brand-self identification |

## Conclusion

The present research was conducted for providing a systematic overview of BL to present its history and future prospective. Using advanced bibliometric techniques, the researchers created a database of main complementary studies based on Scopus. By applying filters, 1,468 documents were accordingly made available to the researchers. They examined a number of performance components such as author, publication, country, and university. After plotting scientific maps, the researchers specified the roadmap for future studies to determine intellectual structure in BL. The trend of publications on BL was first reviewed and it was then found that this domain continues to grow, being of interest to researchers. Influential articles in BL with highest number of citations were then identified, being addressed in highly cited articles on brand love, brand special value, and social media. Bagizi and Wang were also found influential authors. The publications related to this domain based on total number of citations and h-index were assessed. It was found that the Journal of Business Research received the most citations with a high h-index in BL. In the following, the most effective organizations and countries within this domain are introduced. Results showed that USA had both the most production and citations followed by Asian countries with high production in BL, indicating that it is highly considered by researchers in Asia so that in the review of institutions, South Korean and Australian universities had larger numbers of productions.

In network analysis and scientific maps section, the co-word map with four clusters was plotted. The most frequent words were then "BL", "brand value", and "brand trust". Collaboration between authors showed that the cooperation of countries producing articles within the field of BL was steady between authors from USA and South Korea. These two countries have remarkable impacts on this field, while there is scattered collaboration between authors in this domain. The co-authorship network also showed that Keller, Hair, and Aaker had the highest co-citation in BL domain, respectively. In the meantime, the highest number of citations in BL articles was related to works by Keller with 1,173 citations. Finally, by analyzing keywords in BL in both forms of WordCloud and word growth, it was found that social media and brand experience are of interest for researchers in this domain, being observed in both WordClouds. Therefore, further growth of articles in BL centered on social media will be observed in coming years. Review of keywords also showed that there has been no work in this domain related to COVID-19 pandemic. Analyzing highly cited articles over the last four years (mostly addressing tourism and hoteling industries) using covariance-based software indicated that their methodologies were quantitative. Influential factors and the most frequent outcomes of BL were also recognized in these articles. The paper contributed to the literature by providing a comprehensive and updated overview of BL research, and by highlighting the opportunities and challenges for future studies.

## Limitations and suggestions for future research

Referring to only one single scientific database, and using English-only articles could be viewed as limitations of the present work. Future research is therefore suggested to use papers written in other languages related to BL as well as other sources like books and conference papers. Nevertheless, the researchers attempted to make a precise analysis of scientific documents through comprehensive review of BL so that their findings could be beneficial for stakeholders and research communities as a primary reference for gaining insight into BL. It is suggested that further studies should explore the impact of social media and brand experience on BL, as well as the effect of COVID-19 pandemic on BL in different industries.